\documentclass[twocolumn,showpacs,prl,floatfix]{revtex4}

\usepackage[pdftex]{graphicx}
\usepackage{amsmath}
\usepackage{amssymb}
\usepackage{lipsum}
\usepackage{color}
\usepackage{bm}

\newcommand{\fv}[1]{\left\langle #1 \right\rangle}

\begin{document}
\title{Mesoscale model of dislocation motion and crystal plasticity}

\author{Audun Skaugen\footnote{audun.skaugen@fys.uio.no} and Luiza Angheluta}
\affiliation{Njord Center, Department of Physics, University of Oslo, P.O. 1048 Blindern, 0316 Oslo, Norway}
\author{Jorge Vi\~nals}
\affiliation{School of Physics and Astronomy, University of Minnesota, 116 Church St. SE, Minneapolis, MN 55455, USA}

\begin{abstract}
A consistent, small scale description of plastic motion in a crystalline solid is presented based on a phase field description. By allowing for independent mass motion given by the phase field, and lattice distortion, the solid can remain in mechanical equilibrium on the timescale of plastic motion. Singular (incompatible) strains are determined by the phase field, to which smooth distortions are added to satisfy mechanical equilibrium. A numerical implementation of the model is presented, and used to study a benchmark problem: the motion of an edge dislocation dipole in a hexagonal lattice. The time dependence of the dipole separation agrees with classical elasticity without any adjustable parameters.
\end{abstract}
\date{\today}


\maketitle

A phase field theory of a crystalline solid is one of the contending approaches to model defect mediated plastic motion at the nanoscale \cite{re:elder02,re:emmerich12}. Due to the diffusive nature of the evolution of the phase field, existing formulations do not maintain mechanical equilibrium during defect motion, and hence are not adequate models of plastic motion. We reinterpret the phase field as the source for singular strains in defected configurations, and add smooth distortions to maintain mechanical equilibrium at all times. The relative motion of two edge dislocations of opposite Burger's vectors under each other's stress field is studied in a hexagonal lattice. In contrast with a direct solution of the phase field equations, we recover the classical law in which the dipole separation scales with time as $\sqrt{t_0 - t}$.

Crystal plasticity theory assumes coarse grained volumes that contain a large number of defects, and hence is valid at scales on the order of microns or larger. On the other hand, recent theoretical efforts \cite{re:acharya01,re:elder02,re:haataja02,re:koslowski02,re:elder04,re:limkumnerd06,elder2007phase,re:acharya12,re:acharya15,re:groma15} focus on the nanoscale, as new high resolution experiments and large scale simulations are producing a wealth of information about defect motion at the atomic scale. State of the art Bragg Coherent Diffractive Imaging can determine atomic scale displacements with $\leq 30$~nm resolution \cite{re:rollett17,re:suter17}. Advanced image processing methods allow the determination of the strain field phase around a single defect, clearly evidencing its multivalued nature. Indeed, single dislocations have been successfully imaged and their motion tracked quantitatively just recently \cite{re:yau17}. At the same time, nanoscale experiments are revealing new phenomenology that shows that plastic distortion is a rather complex process, characterized by strain bursts and dislocation avalanches \cite{Weiss2003,re:koslowski04,re:csikor07,re:ispanovity10,Mikko2014,re:tarp14,re:cui16}. These phenomena lie squarely outside of classical crystal plasticity theory. 

A continuum theory of plasticity starts from the statement of incompatibility of the deformation gradient tensor
\begin{equation}
\epsilon_{ilm} \partial_{l} w_{mk} = \alpha_{ik}, \quad
w_{mk} = \partial_{m} u_{k}
\label{eq:dislocation_density}
\end{equation}
where $\epsilon_{ilm}$ is the anti symmetric Levi-Civita tensor, $\alpha_{ik}$ the dislocation density tensor, and $w_{mk}$ the distortion tensor \cite{re:kosevich79,re:nelson81}. The integral of $\alpha_{ik}$ over a surface is the sum of the Burger's vectors $\mathbf b$ corresponding to all the dislocation lines $n$ that pierce the surface $\int_{S} \alpha_{ij}dS_{j} = \sum_{n} b_{i}^{n}$. For any given distribution of topological defects in the material, $\alpha$ is fixed, but not the distortion, which can be decomposed into a singular part, the curl of which yields $\alpha_{ij}$, and a smooth strain which we denote by $u_{ij}^{\delta}$. The smooth strain is compatible $ \epsilon_{ikm}\epsilon_{jln}\partial_{kl}u_{mn}^{\delta} = 0$. Regardless of the state of distortion, plastic motion is slow on the scale of lattice vibration, and occurs in mechanical equilibrium, as the stress $\sigma_{ij}$ adiabatically follows the instantaneous distribution of dislocations, $\partial_{j} \sigma_{ij} = 0$. Closure generally requires a constitutive relation involving the stress and the smooth deformation. These considerations and appropriate boundary conditions are sufficient to specify the static problem. Dynamically, over the time scale appropriate for plastic flow, an evolution equation needs to be introduced for the dislocation density tensor \cite{re:rickman97,re:aguenaou97,re:acharya01,re:haataja02,re:limkumnerd06}. In field dislocation dynamics theories, its evolution is kinematically related to the velocity of the dislocation lines, which in turn require a constitutive definition in terms of a local free energy and a dissipation function \cite{re:acharya04}. The phase field crystal model of defect motion, as currently formulated \cite{re:elder02,re:elder04,re:berry06,re:heinonen16}, can be used to specify most of the static and dynamic features just described, but not all, as discussed below.

The phase field, a scalar function of space and time, is a physical order parameter that describes the dimensionless mass density of the crystalline phase $\psi(\mathbf{r}) = \psi_{0} + \sum_{\mathbf{g}} A_{\mathbf{g}} e^{i \mathbf{g} \cdot \mathbf{r}}$, where the sum extends over all reciprocal lattice vectors $\mathbf{g}$ of the lattice. A non convex free energy functional for an isothermal system $\mathcal{F}$ is introduced so that its minimizer $\psi^{*}$ has the desired symmetry of the crystalline phase. Lattice constants appear as parameters. In dimensionless units, we use $\mathcal{F}[\psi] = \int d \mathbf{r} f(\psi,\nabla^{2} \psi)$ with  $f(\psi,\nabla^2\psi)= (\mathcal{L}\psi)^2/2+ r^{2} \psi^2/2+\psi^4/4$, and $\mathcal{L} = 1 + \nabla^2$ \cite{re:elder02,elder2007phase,van2009derivation}. The only remaining constant parameter $r$ is the dimensionless distance away from the symmetry breaking bifurcation. For $r > 0$, $\psi^{*} = 0$ is the only stable solution. For $r < 0$, and depending on the conserved spatial average $\psi_0$, $\psi^{*}$ is periodic with wavenumber unity in our dimensionless units, but of various symmetries. For simplicity, we consider here a two dimensional system where the equilibrium configuration is a hexagonal phase with lattice constant $a = 4\pi/\sqrt 3$. The Burger's vector density in 2D is $B_{k} (\mathbf{r}) = \alpha_{3,k}(\mathbf{r}), k=1,2$. Our results, however, can be readily extended to three dimensions. The temporal evolution of the phase field $\psi$ is defined to be relaxational and driven by free energy reduction 
\begin{equation}
\partial_{t} \psi(\mathbf{r},t) = \nabla^{2} \frac{\delta \mathcal{F}}{ \delta \psi(\mathbf{r},t)}
\label{eq:TDGL}
\end{equation}
where $\delta / \delta \psi(\mathbf{r},t)$ stands for the variational derivative with respect to the phase field. 

For smooth distortions of $\psi^{*}$, the free energy $\mathcal{F}$ suffices to determine the stress-strain relation \cite{re:elder04}. For small distortions, we define a non-singular stress $\sigma^{\psi}$ \cite{skaugen2018dislocation}
\begin{equation}
\label{eq:tildesigmaij}
\sigma_{ij}^{\psi} = \fv{\tilde\sigma^{\psi}_{ij}}_c, \quad
\tilde\sigma_{ij}^{\psi} = \left[\partial_i\mathcal{L}\psi\right]\partial_j\psi - \left[\mathcal{L}\psi\right]\partial_{ij}\psi + f \delta_{ij},
\end{equation}
with $\tilde{\sigma}^{\psi}_{ij}$ given by the local variation of $\mathcal F$ with $\partial_i u_j$, and $\fv{\cdot}_c$ denoting a spatial average across a region roughly corresponding to a unit cell. $\sigma_{ij}^{\psi}$ is symmetric and related to the strain field $u_{ij}= (\partial_i u_j+\partial_j u_i)/2$ according to linear elasticity. For the hexagonal phase under discussion, the relation is that of isotropic elasticity
\begin{equation}
\label{eq:elastic}
 \sigma_{ij}^{\psi} = \lambda \delta_{ij}u_{kk} + 2\mu u_{ij}
\end{equation}
with Lam\'e coefficients $\lambda = \mu = 3 A_0^2$ \cite{skaugen2018dislocation}. The quantity $A_{0}$ is the amplitude of the uniform mode in a multiple scale amplitude expansion of $\psi^{*}$.

The phase field model has been used extensively to describe static and dynamic defected configurations. Following early work on dislocation motion and grain boundaries in roll patterns \cite{re:siggia81,re:tesauro87}, the phase field crystal theory has been used to study dislocation \cite{re:limkumnerd06,re:berry06} and grain boundary motion \cite{re:adland13,re:taha17}. Strain fields have been explicitly extracted \cite{re:schwalbach13}, or imposed to analyze strained film epitaxy \cite{re:huang08}, and considered as the limiting case of phonon degrees of freedom \cite{re:heinonen16}. More complex properties of defect motion such as specification of slip systems, defect mobilities, and Peierls barriers are also given by phase field kinetics \cite{re:boyer02,re:perreault16,skaugen2018dislocation} thus opening the door to the study of defect pinning, bursts, and avalanches.
However, whereas for a specified and fixed defected configuration minimizers of $\mathcal{F}$ with appropriate boundary conditions can be found that are in mechanical equilibrium, any local deformation of $\psi(\mathbf{r},t)$ propagates only diffusively according to Eq.~(\ref{eq:TDGL}). The relevant transverse diffusion constant is small, and can even vanish \cite{re:cross95a}. This is not physical for a crystalline solid, as has been already recognized \cite{re:berry06,heinonen2014phase,re:heinonen16,stefanovic-elastic}. In ordinary crystals, unlike the phase field model, elastic equilibrium compatible with a transient distribution $\alpha_{ik}(\mathbf{r},t)$ and boundary conditions is established quickly, in a time scale determined by damping of elastic waves in the medium.

To overcome this difficulty, we propose to use the phase field $\psi(\mathbf{r},t)$ only as an indicator function of defect location and topology, as well as governing local relaxation near defect cores. The field $\psi(\mathbf{r},t)$ determines the source for lattice incompatibility in Eq.~(\ref{eq:dislocation_density}), the solution of which is only a {\em particular singular solution} for the deformation field. A smooth distortion $\mathbf u^\delta$ (in the null space of the curl) must be added to this particular solution to enforce elastic equilibrium. Equation (\ref{eq:TDGL}) for the newly displaced phase field $\psi^{\prime}(\mathbf{r}) = \psi(\mathbf{r} + \mathbf{u}^{\delta})$ provides for defect motion in a manner that is consistent with the Peach-Kohler force \cite{skaugen2018dislocation}. Plastic motion is uniquely specified, with the only constitutive input being the free energy functional $\mathcal{F}$. We discuss in what follows the details of our computational implementation, and specifically address the relative motion of a dislocation dipole in a 2D hexagonal phase. 

\begin{figure}
\includegraphics[width=0.5\textwidth]{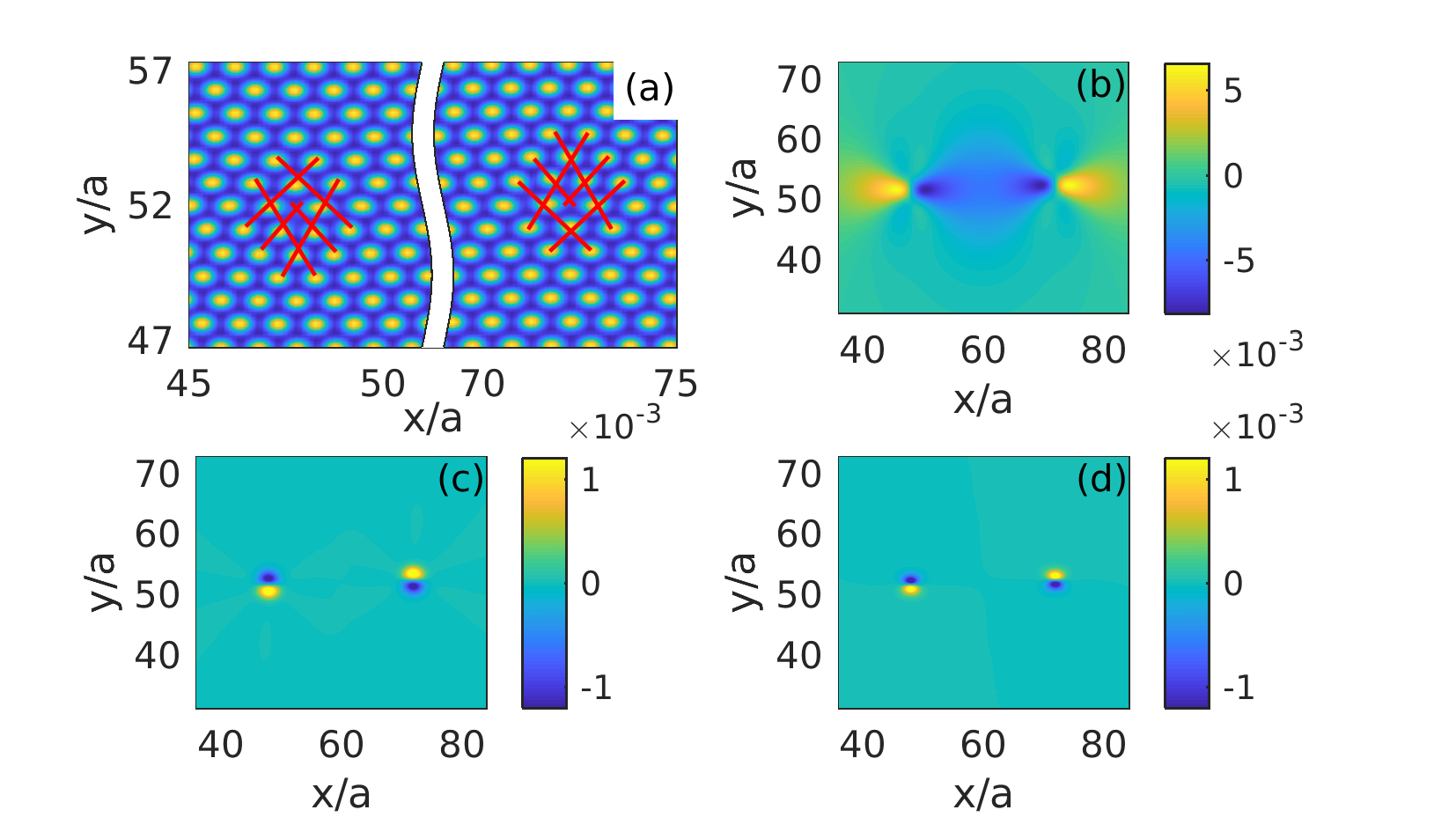}
\caption{(a): Phase field $\psi$ for an initial condition comprising two dislocations with opposite Burger's vectors on the same glide plane. Crystal planes in the $[11]$ and $[\overline{1}1]$ directions are indicated to illustrate the structure of the dislocations in the hexagonal lattice. (b): Coarse-grained shear stress $\sigma_{xy}^{\psi}$ showing two characteristic stress dipoles. (c): Right hand side of Eq.~(\ref{eq:biharm}), divided by $2\mu$, showing dipolar sources at the dislocation positions. (d): Curl of the Burger's vector density as computed by demodulation in ref.~\cite{skaugen2018dislocation}, showing good agreement with (c).}
\label{fig:1}
\end{figure}

We decompose the stress field into a singular part arising from the current phase field configuration, $\sigma^{\psi}$, and a small contribution arising from the smooth distortion $\sigma^{\delta}$, so that $\sigma = \sigma^{\psi}+\sigma^{\delta}$ is in mechanical equilibrium $\nabla \cdot \sigma = 0$. This condition is satisfied by introducing the Airy function $\chi$, which in two dimensions reads $\sigma_{ij}  = \epsilon_{ik}\epsilon_{jl}\partial_{kl}\chi$. Inverting Eq.~(\ref{eq:elastic}), we have in 2D
\begin{equation}
u_{ij} = \frac 1 2 (\partial_i u_j + \partial_j u_i)
= \frac{1}{2\mu}\left(\sigma_{ij} - \kappa\delta_{ij}\sigma_{kk}\right),
\label{eq:lin_elast}
\end{equation}
where $\kappa = \frac{\lambda} {2(\lambda+\mu)}$. Inserting Eq.~(\ref{eq:lin_elast}) into the incompatibility relation in 2D $\epsilon_{ik}\epsilon_{jl}\partial_{kl}u_{ij}=\epsilon_{ij} \partial_{i} B_{j}(\mathbf{r})$ (e.g. \cite{re:kosevich79}) and expressing the stress in terms of $\chi$-function, we obtain that 
\begin{equation}
\frac{1-\kappa}{2\mu} \nabla^4\chi = \epsilon_{ij} \partial_{i} B_{j}(\mathbf{r}),
\label{eq:biharm_sing}
\end{equation}
where $B_j(\mathbf r) = \sum_\alpha b^n_{j} \delta(\mathbf r - \mathbf{r}_n)$ is the dislocation density in 2D for a configuration of dislocations with Burgers vector $\mathbf{b}^n$ at locations $\mathbf r_{n}$. In Ref.~\cite{skaugen2018dislocation}, we explicitly computed $\mathbf{B}(\mathbf{r})$ through complex demodulation of the phase field $\psi(\mathbf{r},t)$. Demodulation yields both the amplitude and phase of the deformation field; the former going to zero at the defect core, the latter undergoing a discontinuity across a line that terminates at the core. Figure \ref{fig:1}(a) shows a dislocation dipole in a 2D hexagonal lattice, and Figure \ref{fig:1}(d) the right hand side of Eq.~(\ref{eq:biharm_sing})) obtained by demodulation. We proceed differently here and introduce a more efficient numerical procedure that does not require demodulation. The smooth strain $u_{ij}^{\delta}$ is compatible ($\epsilon_{ik}\epsilon_{jl} \partial_{ij}u_{kl}^{\delta} = 0$) and therefore, with Eq.~(\ref{eq:lin_elast}), the corresponding stress satisfies,
\begin{equation}
\epsilon_{ik}\epsilon_{jl}\partial_{ij}\left(\sigma_{kl}^\delta - \kappa\delta_{kl}\sigma_{ll}^\delta\right) = 0.
\end{equation}
We now proceed as if the linear decomposition $\sigma = \sigma^{\psi} + \sigma^{\delta}$ holds everywhere, including near dislocation cores as defined by the phase field. This is not strictly correct, but this decomposition results in a distortion $u_{ij}^{\delta}$ which is everywhere compatible. However, as discussed below, the computed stress field $\sigma$ will be divergence free only away from any defect core. Given this decomposition $\sigma_{ij}^{\delta} = \epsilon_{ik}\epsilon_{jl}\partial_{ij}\chi - \sigma_{ij}^\psi$, we find an analogous result to  Eq.~(\ref{eq:biharm_sing}),
\begin{equation}
(1-\kappa)\nabla^4\chi = \left(\epsilon_{ik}\epsilon_{jl}\partial_{ij}\sigma_{kl}^{\psi} - \kappa\nabla^2\sigma_{kk}^{\psi} \right). \label{eq:biharm}
\end{equation}
Note that the stress $\sigma^\psi$ (as defined in Eq.~(\ref{eq:tildesigmaij})) is smooth and bounded, so the right-hand side of Eq.~(\ref{eq:biharm}) can only give a nonsingular approximation to the singular right hand side of Eq.~(\ref{eq:biharm_sing}). Figure \ref{fig:1}(c) shows the right-hand side of Eq.~(\ref{eq:biharm}) obtained numerically for the dislocation dipole which is in good agreement Eq.~(\ref{eq:biharm_sing}) obtained through demodulation (Fig.~\ref{fig:1}(d)). Both methods act as regularizations of the singular density at defect cores.

From a given phase field configuration at time $t$, $\psi(\mathbf{r},t)$, we compute $\sigma^{\psi}$ from Eq.~(\ref{eq:tildesigmaij}), and then solve Eq.~(\ref{eq:biharm}) to obtain $\chi$ and therefore $\sigma$. The difference $\sigma_{ij}^{\delta} = \epsilon_{ik}\epsilon_{jl} \partial_{ij} \chi - \sigma_{ij}^{\delta}$ leads to the smooth strain $u_{ij}^{\delta} = \frac{1}{2 \mu} ( \sigma_{ij}^{\delta} - \nu \sigma_{kk}^{\delta} \delta_{ij} )$ which is, by construction, compatible. It can, therefore, be integrated to obtain a compatible deformation $\mathbf{u}^{\delta}$. The final step in the computation is to redefine the phase field $\psi^{\prime} (\mathbf{r},t) = \psi(\mathbf{r} + \mathbf u^\delta,t)$. 

\begin{figure}
\includegraphics[width=0.5\textwidth]{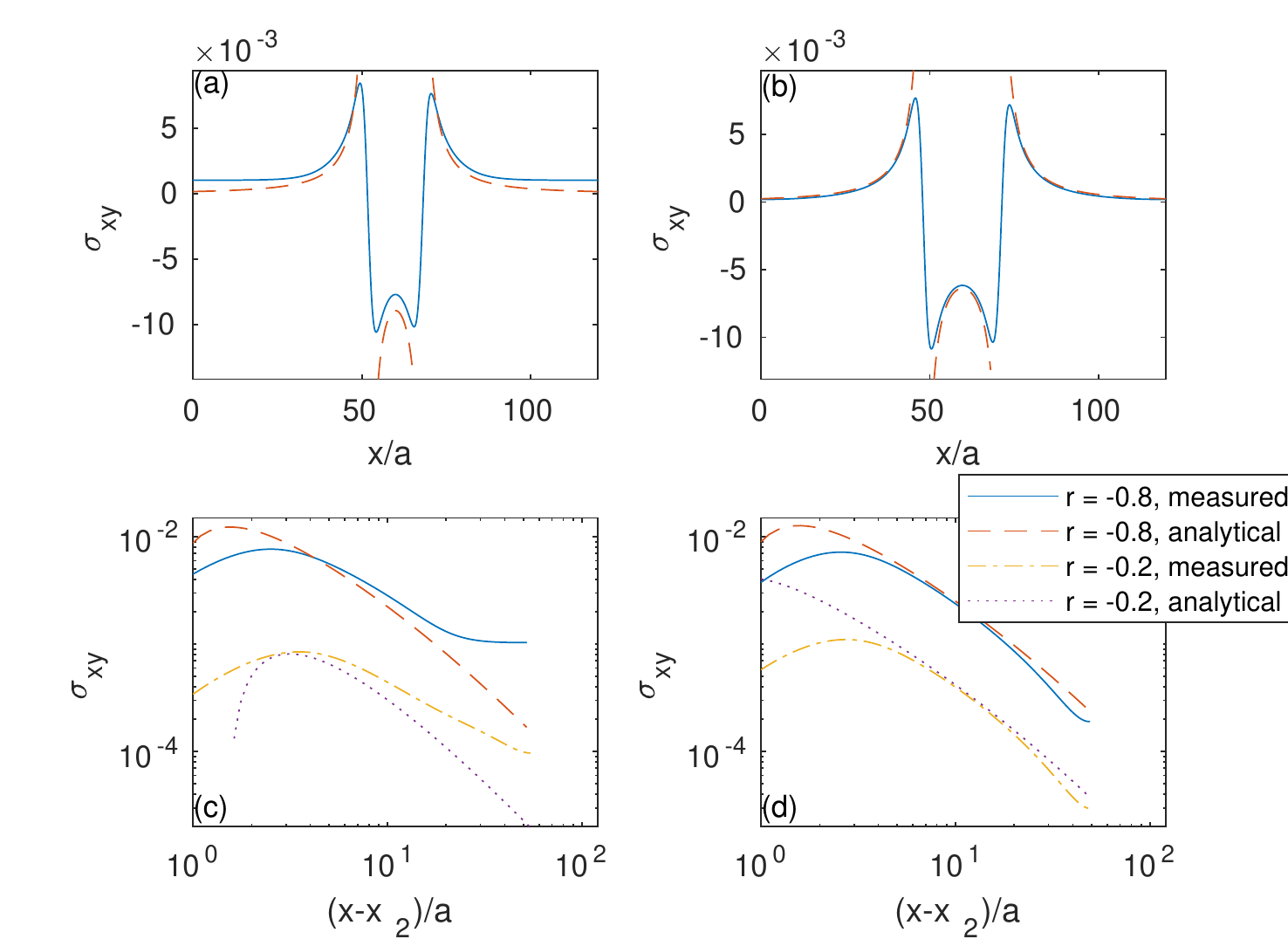}
\caption{(a) Shear stress $\sigma^\psi_{xy}$ along the line joining the two dislocation cores from direct integration of Eq.~(\ref{eq:TDGL}) (solid line) compared with Eq.~(\ref{eq:anastress}) (dashed line). (b) Constrained stress according to the simultaneous solution of Eqs.~(\ref{eq:TDGL}), (\ref{eq:biharm}), and (\ref{eq:final_stress}), also compared with Eq.~(\ref{eq:anastress}). (c) and (d) are the enlarged tail regions of (a) and (b) respectively.}
\label{fig:2}
\end{figure}

Both the stress-strain relation and stress superposition only hold far from defect cores. We define the stress of this newly deformed configuration everywhere as
\begin{equation}
\sigma_{ij} = \sigma^\psi_{ij} + \sigma^\delta_{ij} = \sigma_{ij}^{\psi} + \lambda \delta_{ij}u^\delta_{kk} + 2\mu u^\delta_{ij},
\label{eq:final_stress}
\end{equation}
which satisfies $\partial_{j} \sigma_{ij} = 0$ only far from defect cores, not at short distances. This is not a problem as standard diffusive evolution of the phase field suffices to equilibrate the stress near cores in time. We discuss this further below, and in Fig.~\ref{fig:2}.

The integration of the compatible strain $u_{ij}^\delta$ to obtain $\mathbf u^\delta$ is carried out through a Helmholtz decomposition into curl-free and divergence-free parts $ u_i^\delta = \partial_i V + \epsilon_{ij}\partial_j A$. Applying the divergence to this expression, one obtains a Poisson equation for the potential $V$, $\partial_i u_i^\delta = u_{ii}^\delta = \nabla^2 V,$ which is easily solved by spectral methods. On the other hand, taking the curl we find 
$\epsilon_{ij}\partial_i u_j^\delta = \epsilon_{ij}\epsilon_{jk}\partial_{ik}A = -\nabla^2 A,$ which is a Poisson equation for $A$. Unfortunately the source term depends on the antisymmetric part of the smooth deformation gradient, which we do not obtain directly from the elastic stress, as this only depends on the symmetric part. We therefore apply another Laplacian operator to the equation, and use the compatibility relation $\epsilon_{ij}\partial_{ij}u^\delta_k = 0$ to find
\begin{equation}
\nabla^4 A = -\epsilon_{ij}\partial_{ik}(\partial_k u_j^\delta + \partial_j u_k^\delta)
= -2\epsilon_{ij}\partial_{ik}u_{jk}^\delta. 
\end{equation}
This is a biharmonic equation for $A$ with a known source term, which is again easily solved by spectral methods. In particular, if $k_i$ are the components of the $\mathbf k$ vector and $\hat{u}_{ij}^\delta$ are the Fourier components of the residual strain, the Fourier components of the residual deformation can be expressed as 
\begin{equation}
\hat{u}_i^\delta = - \frac{ik_i}{k^2}\hat{u}_{jj}^\delta + 2i\epsilon_{ij}\epsilon_{rs}\frac{k_jk_rk_l}{k^4}\hat{u}_{sl}^\delta,
\label{eq:smooth_distortion}
\end{equation}
with the $k=0$ component chosen to be zero to avoid rigid body displacements. We then compute the distorted phase field $\psi'(\mathbf r) = \psi(\mathbf r + \mathbf u^\delta)$ on the original grid $\mathbf r$ by expanding in Taylor series up to fifth order in $\mathbf{u}^{\delta}$.

We present next the results of a numerical study of a benchmark configuration: the relative motion of two edge dislocations along their glide plane, Fig.~\ref{fig:1}(a). The computational domain is assumed periodic containing 120x120 unit cells with a spatial resolution of $a/8 = \pi/2\sqrt 3$ in the $x$ direction, and $2\pi/7$ in the $y$ direction. The initial distance between dislocations is $24a$, and we consider two sets of values $r = -0.2, \psi_0 = 0.265$ and $r = -0.8,\psi_0 = -0.43$. We prepare the initial condition in the same way as in Ref.~\cite{skaugen2018dislocation}, and numerically solve Eq.~(\ref{eq:TDGL}) using an exponential time differencing method with a time step of $\Delta t = 0.1$ \cite{cox2002exponential}. According to linear elasticity theory for an isotropic medium, the shear stress in such a configuration is,
\begin{equation}
\sigma_{xy} = \frac{2\mu(\lambda+\mu)}{\lambda+2\mu} \sum_\alpha \frac{b_x^n}{2\pi}\frac{\cos\phi_n \cos(2\phi_n)}{|\mathbf r-\mathbf r_n|},
\label{eq:anastress}
\end{equation}
where $\phi_n$ is the azimuth relative to dislocation $n$. Figures (\ref{fig:2})(a,c) show $\sigma^{\psi}$ along a line that includes the two dislocation cores at time $t = 700$ obtained by direct integration of Eq.~(\ref{eq:TDGL}), and compares it to Eq.~(\ref{eq:anastress}). Divergences in Eq.~(\ref{eq:anastress}) are regularized by the phase field, and the stress near the cores is relatively well described by $\sigma^{\psi}$. Far from the cores, however, the two stresses show qualitatively different asymptotic dependence. Figures (\ref{fig:2})(b,d) show the stress in a configuration in which the smooth distortion (\ref{eq:smooth_distortion}) has been applied between time steps. The stress is still regularized near defect cores, yet, by construction, agrees with linear elasticity away from them.

\begin{figure}
\includegraphics[width=0.5\textwidth]{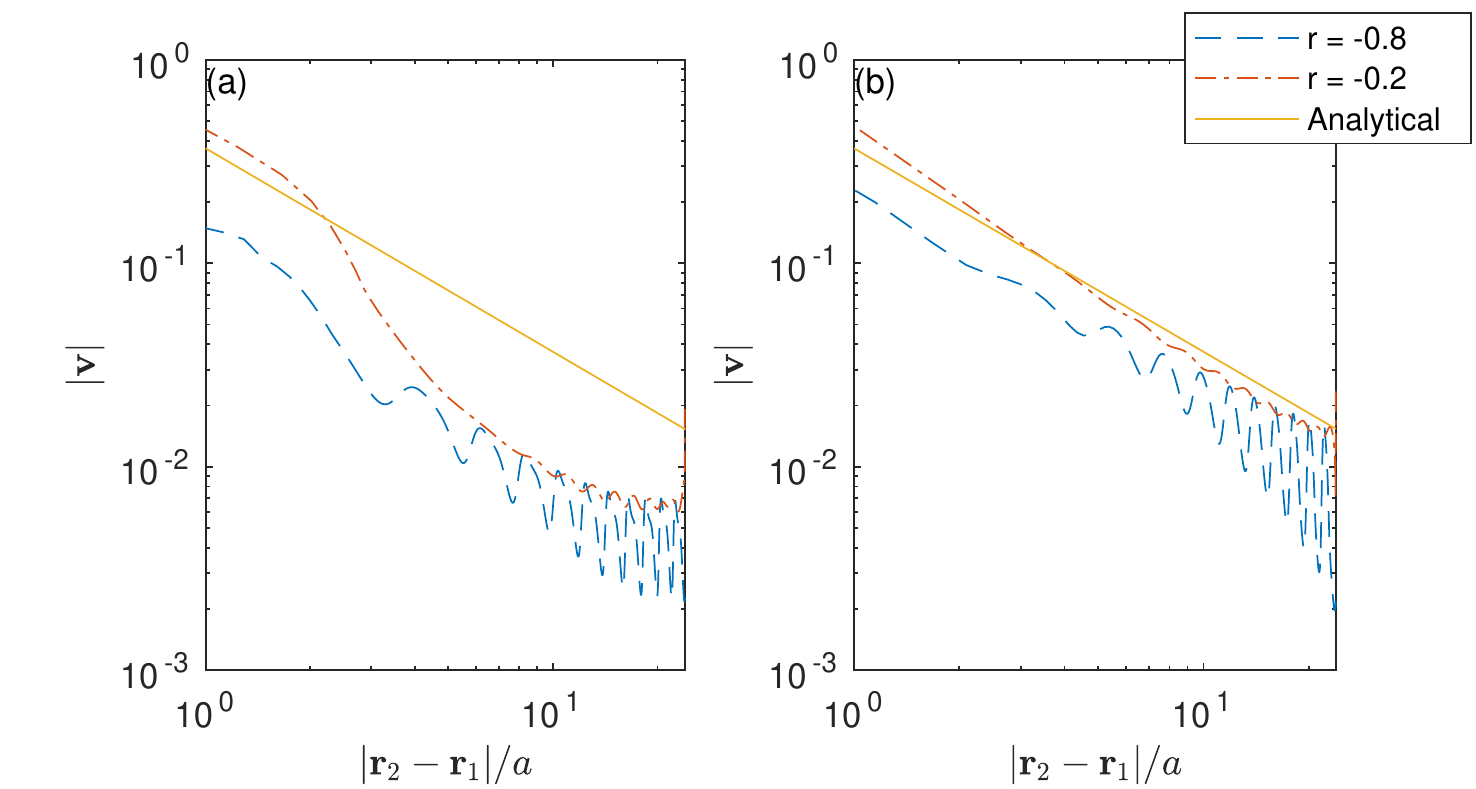}
\caption{Dislocation velocity as a function of dipole separation. Velocities from the numerical computations are obtained from the locations of the zeros of the complex amplitudes of $\psi$ as described in Ref.~\cite{skaugen2018dislocation}. The analytic result follows from the Peach-Kohler force with stress given by Eq.~(\ref{eq:anastress}), and mobility computed from $\mathcal{F}$ as given in \cite{skaugen2018dislocation}. There are no adjustable parameters in this calculation. (a), Numerically computed velocity by direct integration of Eq.~(\ref{eq:TDGL}). (b), Velocity given by our model.}
\label{fig:3}
\end{figure}

Figure \ref{fig:3} shows the time dependence of the dislocation velocity as a function of dipole separation as given by direct integration of the phase field model, Eq.~(\ref{eq:TDGL}), Fig.~\ref{fig:3}(a), and by our model, Fig.~\ref{fig:3}(b). For reference, we also show the expected result from elasticity theory by using the Peach-Kohler force with stress (\ref{eq:anastress}), and mobility derived from $\mathcal{F}$ (Eq.~(45) in Ref.~\cite{skaugen2018dislocation}). There are no adjustable parameters in the calculation of the analytic velocity. The two dislocations move towards each other until they annihilate, with a velocity inversely proportional to their separation. Our model captures this result well for a range of parameters $r$, with slight stick-slip motion observed at larger $|r|$, visible as oscillations in the dislocation velocity.

To summarize, we have argued that the phase field crystal model currently in use lacks deformation as an independent variable, and as a consequence fails to maintain proper mechanical equilibrium during plastic motion. We retain the model because it provides for lattice and topological defect structures as derived properties from the phenomenological free energy. It also allows regularization of defect cores and singular stresses. Phase field kinetics is also consistent with the classical Peach-Kohler force, with mobility that is again specified by the free energy $\mathcal{F}$. We take the view, however, that the phase field is not adequate to describe the distortion of the lattice away from moving defect cores, and hence supplement it with a smooth distortion field, compatible with the topological content of the phase field, but defined so as to maintain mechanical equilibrium everywhere away from defect cores. When the evolution of $\psi'(\mathbf r,t)$ is thus constrained to satisfy mechanical equilibrium at all times, we show numerically that our model agrees with the classical law of motion for a dislocation dipole in isotropic, linear elasticity. These results put the phase field crystal model on firmer ground to study more complex defected configurations at the nanoscale.

\acknowledgments{This research has been supported by a startup grant from the University of Oslo, and the National Science Foundation under contract DMS 1435372.}

\bibliographystyle{apsrev4-1}
\bibliography{ref}

\end{document}